\documentclass[english,journal=jctcce,manuscript=review]{achemso}
\usepackage[T1]{fontenc}
\usepackage[latin9]{inputenc}
\usepackage{color}
\usepackage{amstext}
\PassOptionsToPackage{version=3}{mhchem}
\usepackage{mhchem}

\makeatletter

\title{Transition moments for STEOM-CCSD with core triples}

\author{Megan Simons}
\email{msimons@smu.edu}

\author{Devin A. Matthews}

\affiliation{Department of Chemistry, Southern Methodist University, Dallas, TX
75275, USA}

\usepackage{mhchem}
\usepackage[binary-units=true]{siunitx}

\makeatother

\usepackage{babel}
\begin{document}
\begin{tocentry}
\includegraphics{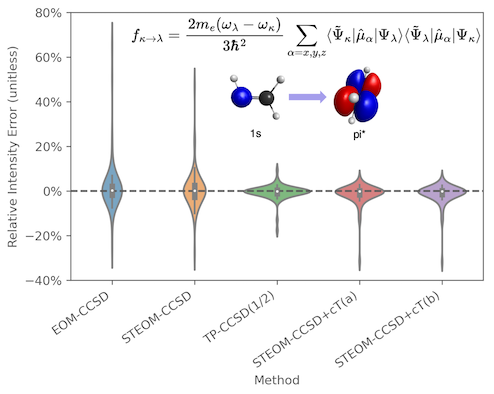}
\end{tocentry}
\begin{abstract}
Similarity transformed equation-of-motion coupled cluster theory (STEOM-CC) is an alternative approach to equation-of-motion coupled cluster theory for excited states (EOMEE-CC) which uses a second similarity transformation of the Hamiltonian, followed by diagonalization in a small (CI singles-like) excitation space, even when single and double excitations are included in the transformation. In addition to vertical excitation energies, transition moments measure the strength of the interactions between states determining absorption, emission, and other processes. In STEOM-CCSD, transition moments are calculated in a straight-forward manner as biorthogonal expectation values using both the left- and right-hand solutions, with the main difference from EOMEE-CC being the inclusion of the transformation operator. We recently developed an extension of STEOM-CCSD to core excitations, CVS-STEOM-CCSD+cT, which includes triple excitations and the well-known core-valence separation for the core ionization potential calculations. In this work, we derived transition moments for core-excited states with core triple excitations, including both ground-to-core-excited and valence-to-core-excited transitions. The improvement of the computed transition moments of the CVS-STEOM-CCSD+cT method is compared to standard CVS-STEOMEE-CCSD and CVS-EOMEE-CCSD for our previously published small molecule benchmark set.
\end{abstract}

\section{Introduction}

A fundamental objective of electronic structure theory is to describe the properties of the ground state. While this does create a partial picture, it is important to describe the many transition properties for molecules being excited from the ground state to an excited state or from one excited state to another. Excited state energies, and to some extent geometries, are easily benchmarked and can be compared against various experimental or theoretical references. However, this is not the case for other properties, such as oscillator strengths, dipole moments, and vibrational frequencies. \cite{oscstrength2011, worstertransprops} The need for formally-derived transition properties, especially between two excited states, is incredibly important to study experimental spectra and theoretical methods such as time-resolved x-ray spectroscopy, x-ray absorption and emission, and non-adiabatic coupling. In x-ray absorption spectroscopy, computed oscillator strengths aid in the assignment of peaks in experimental spectra.\cite{heideXrayPhotoelectronSpectroscopy2011, normanSimulatingXraySpectroscopies2018} The focus on transition moments allows for insight on the strength of the interactions between states governing absorption, emission, and other processes. Oscillator strengths, $f$, indicate the probability of a given transition from the ground state to an excited state occurring and can be measured through observed intensities.\cite{tddftdipole, oscstrength2011}

Many theoretical chemists utilize density functional theory (DFT) for computational chemistry problems involving the ground state, and time-dependent density functional theory (TD-DFT) extends the concepts of DFT to excited states.\cite{marquestddft,grosstddft} TD-DFT has become popular for calculating excitation energies and excited state properties due to its ability to be accurate, yet computationally efficient.\cite{oscstrength2011, tddftdipole} However, there are self-interaction errors when computing the description of the ground state.\cite{besleytddft, worstertransprops} Other methods used for calculating excited state properties include equation-of-motion coupled cluster (EOM-CC) theory\cite{sekinoLinearResponseCoupledcluster1984,stantonEquationMotionCoupledcluster1993,comeauEquationofmotionCoupledclusterMethod1993} and transition-potential density functional theory (TP-DFT).\cite{stenerDensityFunctionalCalculations1995,huDensityFunctionalComputations1996,trigueroCalculationsNearedgeXrayabsorption1998} Even though EOM-CC, particularly when used with the core-valence separation technique,\cite{corianiCommunicationXrayAbsorption2015} has success in describing valence excitations for core-hole states, there are still large orbital relaxation errors up to 5 eV and there is a large computation cost for each excited state.  TP-DFT is a compromise between linear-response methods and state-specific orbital optimizations, but errors remain when assigning peak positions and intensities in x-ray absorption spectra.\cite{huDensityFunctionalComputations1996, trigueroCalculationsNearedgeXrayabsorption1998,trigueroSeparateStateVs1999}
We recently introduced a transition-potential coupled cluster (TP-CC) method that combines the concepts of coupled cluster theory and TP-DFT in order to reduce the orbital relaxation error present in EOM-CCSD. \cite{simonsTransitionpotentialCoupledCluster2021, simonsopttpcc} We also explored similarity-transformed equation of motion coupled cluster (STEOM-CC) theory, (originally proposed by Nooijen and Bartlett \cite{nooijenNewMethodExcited1997,NooijenSimilaritytransformedequationofmotion1997}) an alternative to EOM-CC theory, for excitation energies for XAS and introduced a CVS-STEOM-CCSD+cT method that includes triple excitations only in the core ionized potential to account for the orbital relaxation present in STEOMEE-CC methods.\cite{simonssteom}

In this paper we present a comparison of oscillator strengths for CVS-EOM-CCSDT, TP-CCSD(1/2), and CVS-(ST)EOM-CCSD methods. Additionally, we present the implementation and calculation of oscillator strengths for CVS-STEOM-CCSD+cT, which has an explicit triples contribution only for the core orbital(s).

\section{Theoretical Methods}

The details of the CVS-STEOMEE-CCSD+cT method are available in our previous publication,\cite{simonssteom} while the theory of transition moments in similarity-transformed equation-of-motion coupled cluster was developed by Nooijen in his original publications.\cite{NooijenSimilaritytransformedequationofmotion1997} Here we briefly recap the important features of the core triples and their effect on the computed STEOM transition moments.

In STEOMEE-CCSD+cT,\cite{simonssteom} the relaxation effects of the core hole are accounted for by including triple excitations in the solution of the core ionization potential equations via CVS-EOMIP-CCSDT.\cite{734c7d7e72234a548c61e39e075d7bd2} This results in a three-body transformation operator which is derived from the triples amplitudes of the core EOMIP solution,
\begin{align}
\hat{S}^-_3 &= \frac{1}{12}\sum_{ijkmbc} s^{cbm}_{kji} a_m^\dagger a_b^\dagger a_c^\dagger a_k a_j a_i \\
s^{cbm}_{kji} &= - \sum^{n_{o;act}}_{\kappa\lambda=1} r^{cb}_{kji}(\lambda)(U^{-1}_-)_{\lambda\kappa} \delta_{\kappa m}
\end{align}
where $m$ and at least one of $ijk$ must be an active core orbital. Formally, we consider $r^{cb}_{kji}=0$ for valence or inactive core ionized states. The addition of this transformation operator modified the form of the twice-transformed Hamiltonian, $\hat{G}$, and the resulting STEOM eigenstates. In our implementation we also employ the core-valence separation\cite{corianiCommunicationXrayAbsorption2015} in the diagonalization of the singles-singles block of $\hat{G}$, but the effect of this approximation on the energies should be very small given the lack of coupling to high-lying doubly-excited valence determinants.

As in standard equation-of-motion coupled cluster theory, the oscillator strength for excitation from state $\kappa$ to state $\lambda$ is computed from non-hermitian transition dipole moments \cite{stantonEquationMotionCoupledcluster1993} via an expectation value formalism,
\begin{align}
f_{\kappa \rightarrow \lambda} = \frac{2m_{e}(\omega_\lambda-\omega_\kappa)}{3\hbar^{2}} \sum_{\alpha=x,y,z} \langle \tilde{\Psi}_\kappa|\hat{\mu}_\alpha|\Psi_\lambda\rangle \langle\tilde{\Psi}_\lambda|\hat{\mu}_\alpha|\Psi_\kappa\rangle
\end{align}
Due to the non-hermitian nature of EOM-CC and STEOM-CC, $\langle\tilde{\Psi}_\kappa|$ and $|\Psi_\kappa\rangle$ are distinct. In CVS-STEOMEE-CCSD+cT, these are,
\begin{align}
\langle\tilde{\Psi}_\kappa| &= \langle 0| \hat{L}(\kappa) (1 - \hat{S}_2) e^{-\hat T} \\
|\Psi_\kappa\rangle &= e^{\hat{T}} (1 + \hat{S}_2 + \frac{1}{2} \hat{S}_2^2 + \hat{S}^-_3) \hat{R}(\kappa) |0\rangle
\end{align}
where $|0\rangle$ is the (usually Hartree--Fock) reference determinant. The right-hand ground eigenstate is trivially $\hat{R}(0) = 1$. The left-hand ground eigenstate is formally an eigenvector of $\hat{G}$, but here we use an approximation where the standard left-hand (EOM-)CC eigenstate is used, $\hat{L}(0) = 1 + \hat{\Lambda}$.

The addition of the core triples in CVS-STEOMEE-CCSD+cT then leads to the additional term in the transition dipole moment between states $\kappa$ and $\lambda$,
\begin{align}
\mu_\alpha^{\kappa\lambda}(\mathrm{cT}) &= \langle 0| \hat{L}_2(\kappa) \hat{\mu}_\alpha \hat{S}^-_3 \hat{R}_1(\lambda) |0\rangle = \sum_{ai} D_{ai}^{\kappa\lambda}(\mathrm{cT}) \mu_{ai;\alpha} \label{eq:direct-triples} \\
D_{ai}^{\kappa\lambda}(\mathrm{cT}) &= \langle 0| \hat{L}_2(\kappa) \{ a_a^\dagger a_i \} \hat{S}^-_3 \hat{R}_1(\lambda) |0\rangle \\
&= -\frac{1}{2} \sum_{efmno} l^{ef}_{no}(\kappa) s^{afm}_{ion} r^e_m(\lambda)
-\frac{1}{4} \sum_{efmno} l^{ef}_{no}(\kappa) s^{efm}_{noi} r^a_m(\lambda)
\end{align}
Because the addition of $\hat{S}_3^-$ modifies $\hat{G}$ and hence $\hat{L}(\kappa)$ and $\hat{R}(\kappa)$, the transition moments in CVS-STEOMEE-CCSD+cT are already different from those in CVS-STEOMEE-CCSD without considering \eqref{eq:direct-triples}. Thus, we term the contribution arising directly from the inclusion of $\hat{S}^-_3$ in the transition dipole moment expression in \eqref{eq:direct-triples} as the "direct" triples contribution.

Finally, we note that while $\hat{R}(\kappa)$ consists only of single excitations (and potentially a small contribution from the reference determinant), the left-hand eigenstate $\hat{L}(\kappa)$ formally spans both single and double excitations. The left-hand singles amplitudes may be determined from the singles-singles block of $\hat{G}$ alone, but the doubles amplitudes would require a costly diagonalization in the full singles and double space. Following Nooijen, we use a perturbative approximation for $\hat{L}_2(\kappa)$, although we opt for a simpler approximation which is consistent through first-order,
\begin{align}
l^{ij}_{ab}(\kappa) = \frac{1+P^{ai}_{bj}}{\epsilon_{i}+\epsilon_{j}-\epsilon_{a}-\epsilon_{b} + \omega_\kappa} \left(\sum_{e}l^{i}_{e}(\kappa)\langle ab||ej\rangle -\sum_{m}l^{m}_{a}(\kappa) \langle ij||mb\rangle \right)
\end{align}
where $P^{ai}_{bj}$ exchanges the labels $ai$ and $bj$ in the following expression.

\section{Computational Details}

Transition energy and transition moment calculations for CVS-EOMEE-CCSDT, CVS-EOMEE-CCSD, TP-CCSD(1/2), CVS-STEOMEE-CCSD, and CVS-STEOMEE-CCSD+cT were implemented in a development version of the CFOUR program package.\cite{matthewsCoupledclusterTechniquesComputational2020} A single core orbital was included in the CVS treatment and STEOM principal IP solution in each calculation. 

The test set used and methodology are the same for transition energy and transition moment calculations, as in previous papers.\cite{simonsTransitionpotentialCoupledCluster2021,simonsopttpcc,simonssteom} The test set consisted of all 1s principal core ionizations and four vertical core excitation energies from each 1s core orbital of $\ce{H2O}$, CO, HCN, HF, HOF, HNO, $\ce{CH2}$, $\ce{CH4}$, $\ce{NH3}$, $\ce{H3CF}$ $\ce{H3COH}$, $\ce{H2CO}$, $\ce{H2CNH}$, and $\ce{H2NF}$. The core excitations (and oscillator strengths) were selected as those for which we could reliably converge all methods tested, which typically consisted of the first four excitations of dominant single excitation character. All calculations utilized the aug-cc-pCVTZ basis set with all electrons correlated, except for $\ce{H2O}$ where aug-cc-pCVQZ was used. We have used full CVS-EOM-CCSDT as a benchmark to avoid the complications coming from missing relativistic effects, basis set incompleteness, and geometric effects. \cite{734c7d7e72234a548c61e39e075d7bd2, liuBenchmarkCalculationsKEdge2019} The rationale for choosing the benchmark is the same as in previous work. \cite{simonsTransitionpotentialCoupledCluster2021, simonsopttpcc, simonssteom}

\section{Results and Discussion}

\begin{figure}
    \centering
    \includegraphics[width=\columnwidth]{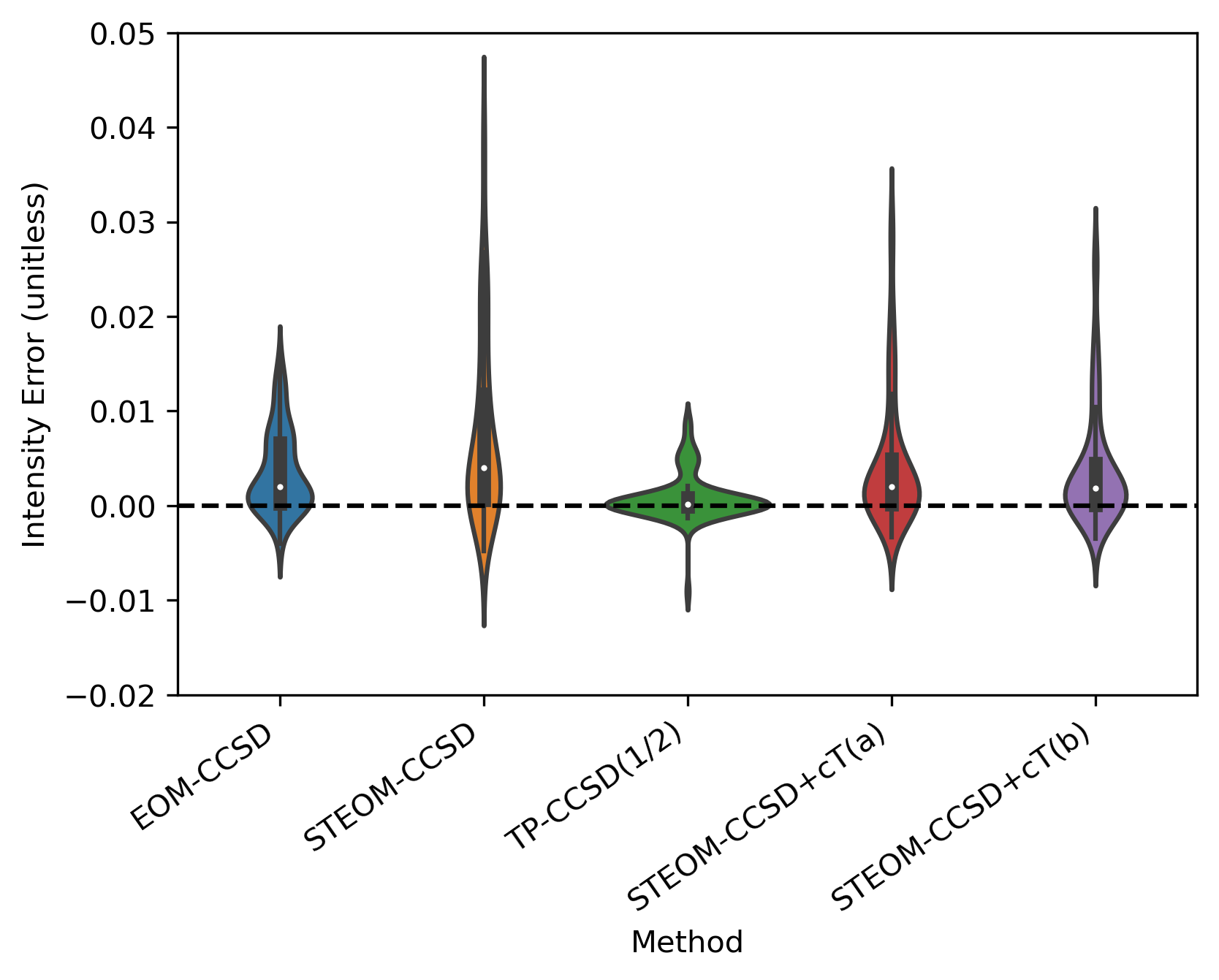}
    \caption{Error distributions with respect to CVS-EOM-CCSDT for absolute oscillator strengths (intensities).}
    \label{fig:interror}
    \raggedright
    \small\textsuperscript{a} Excluding direct core-triples contribution to oscillator strengths. \\
    \small\textsuperscript{b} Including direct core-triples contribution to oscillator strengths.
\end{figure}

\begin{figure}
    \centering
    \includegraphics[width=\columnwidth]{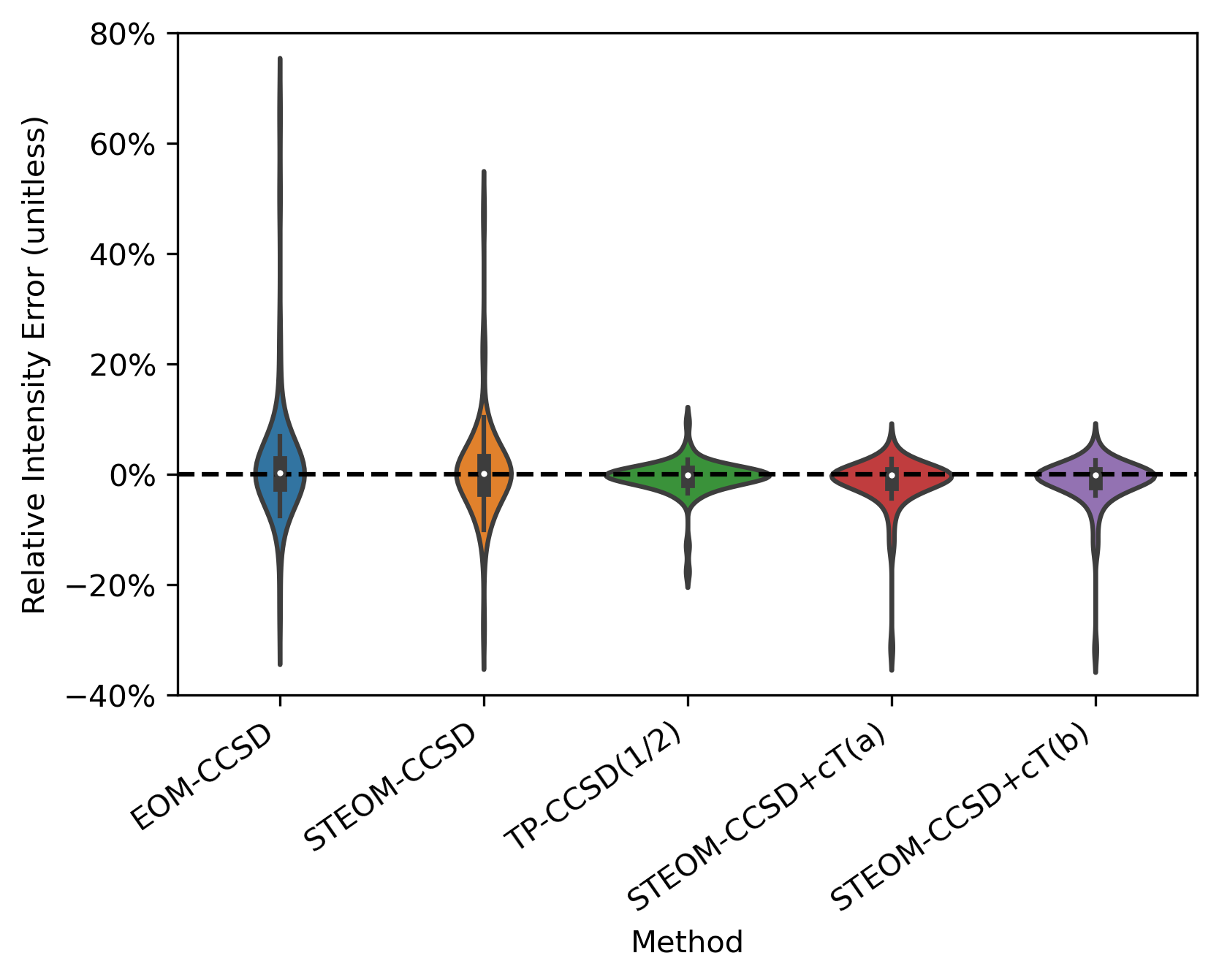}
    \caption{Error distributions with respect to CVS-EOM-CCSDT for relative oscillator strengths (intensities).}
    \label{fig:intrelerror}
    \raggedright
    \small\textsuperscript{a} Excluding direct core-triples contribution to oscillator strengths. \\
    \small\textsuperscript{b} Including direct core-triples contribution to oscillator strengths.
\end{figure}

In the following discussion and in Figures \ref{fig:interror} and \ref{fig:intrelerror}, the "shortened" names of CVS-EOM methods will be used, for example, STEOM-CCSD = CVS-STEOMEE-CCSD, with the exception of TP-CCSD(1/2). The distribution of "absolute" oscillator strength deviations from CCSDT are depicted in Figure \ref{fig:interror}. The absolute oscillator strength deviation for a method $X$ is calculated as $f(X) - f(CCSDT)$. The distribution of "relative" oscillator strength deviations from CCSDT are depicted in Figure \ref{fig:intrelerror}, as percentages. The relative oscillator strength deviation is determined by normalizing each spectrum so that the most intense transition has unit strength.

As can be seen in Fig.~\ref{fig:interror}, the deviations of absolute oscillator strengths indicates an improvement in the TP-CCSD(1/2) and STEOM-CCSD+cT (as measured by standard deviation) over EOM-CCSD and STEOM-CCSD. STEOM-CCSD exhibits a significant increase in standard deviation and outliers (primarily overestimation of oscillator strength) compared to EOM-CCSD. STEOM-CCSD+cT also exhibits a number of outliers compared to TP-CCSD(1/2) and even to EOM-CCSD. All of the STEOM methods have difficulty describing the absolute oscillator strengths of $\pi^*$ and $\sigma^*$ core-to-valence excitated states, with STEOM-CCSD+cT overestimating by an average of $\sim 0.25\times$. Interestingly, the intensity distributions for all methods have a positive skew in comparison with the EOM-CCSDT benchmark. Direct STEOM-CCSD+cT contributions have a slight effect on the absolute oscillator strengths, leading to a $\sim 10\%$ reduction in standard deviation compared to the STEOM-CCSD+cT calculation when direct core triples are excluded. The standard deviation of the error is reduced by almost $40\%$ compared to STEOM-CCSD without any inclusion of core triples.

Normalization of each spectrum with respect to the most intense peak gives us a different viewpoint since all spectra are "equal" while keeping the relative importance of each peak in each spectrum. It can be seen from Fig.~\ref{fig:intrelerror} that both "purely singles and doubles" methods, EOM-CCSD and STEOM-CCSD, are prone to large errors in both the positive and negative direction, as large as $75\%$ of the relative peak intensity. It can also be seen that the STEOM-CCSD+cT methods have essentially the same relative intensity error distributions, indicating an almost complete cancellation of the direct effect of core triples between different peaks in the same spectrum. Nooijen and Bartlett\cite{nooijenNewMethodExcited1997} observed a similar effect for the "triples" contribution arising from $\hat{S}_2^2 \hat{R}_1$, where there was a significant effect on the transition energy but an almost negligible effect on the oscillator strengths (from the direct triples contribution---the indirect effect through the solution of the eigenstates was larger). For TP-CCSD(1/2) and STEOM-CCSD+cT, significant outliers are mostly negative (oscillator strength is underestimated). For STEOM-CCSD+cT, the significant errors are confined almost entirely to the F and C K-edges of \ce{H3CF} and the O K-edge of \ce{H3COH}. Even including these problematic cases, STEOM-CCSD+cT results in a $2.4\times$ reduction in standard deviation compared to EOM-CCSD and over a $1.8\times$ reduction in standard deviation compared to STEOM-CCSD.

The large errors for $\pi^*$ and $\sigma^*$ valence states present in both EOM-CCSD and STEOM-CCSD are almost entirely eliminated in the TP-CCSD(1/2) and STEOM-CCSD+cT, in the latter thanks to a consistent over-estimation of both valence and Rydberg intensities in these spectra. It is not clear why both the valence and Rydberg peaks are so consistently over-estimated, although there could be a consistent error introduced by the left-hand ground state eigenstate used. Also, despite the over-estimation of the valence excitation oscillator strengths, many Rydberg absolute oscillator strengths are not significantly changed from their EOM-CCSD values. There is not enough data yet to indicate if this indicates a link between EOM-CCSD and the over-estimation effects seen in STEOM-CCSD. The overall improvement of the TP-CCSD(1/2) and STEOM-CCSD+cT intensities is due to the improved description of the core-hole, although in differing ways---in TP-CCSD through the fractional occupation in the core-hole which creates a cancellation of errors between the ground and core-excited states, and in STEOM-CCSD+cT through the improved relaxation of the core hole via the inclusion of explicit triple excitations in the core ionization potential cancellation (note that the ionized \emph{wavefunction} is utilized in STEOM, and not just the improved ionization energy). Apart from the cases mentioned above, STEOM-CCSD+cT is seen to estimate the oscillator strengths almost as well as TP-CCSD(1/2).

\section{Conclusions}

Absolute and relative oscillator strengths were calculated for a group of small molecules using various coupled cluster methods, including CVS-EOM-CCSDT, CVS-(ST)EOM-CCSD, TP-CCSD(1/2), and CVS-STEOM-CCSD+cT. Our previous work\cite{simonssteom} showed that the CVS-STEOM-CCSD+cT method performed well for core-excited state energies and was comparable to TP-CCSD(1/2), which has an explicit inclusion of core relaxation via the molecular orbitals. Here, we investigated how well this method estimates transition moments in comparison to full CVS-EOM-CCSDT as the benchmark. CVS-STEOM-CCSD+cT decreases relative errors in oscillators by over $1.8\times$ in comparison to CVS-(ST)EOM-CCSD. All STEOM-CCSD methods are seen to over-estimate the oscillator strengths for $\pi^*$ and $\sigma^*$ core excitations, but do so consistently, leading to an improved ratio of Rydberg intensity to valence intensity. We recommend the use of the CVS-STEOM-CCSD+cT method for core-hole spectroscopy calculations due to its ability to treat core and valence states on an even footing and to accurately predict relative oscillator strengths and transition energies.

\section*{Acknowledgments}

The authors would like to thank Prof. Marcel Nooijen for inspiring us to work on triple excitations in STEOM-CC theory. This work was supported in part by the US National Science Foundation under grant CHE-2143725. MS is supported by an SMU Center for Research Computing Graduate Fellowship. All calculations were performed on the ManeFrame II computing system at SMU.

\section*{Supplementary Material}

An electronic supplementary information file is available as an Excel file (.xslx). This file contains the raw transition energies and oscillator strengths for each orbital K-edge, as well as the absolute and relative transition energies and the absolute and relative oscillator strengths for each orbital K-edge.

\section*{Data Availability}

The data that supports the findings of this study are available within the article and its supplementary material.

\bibliography{paper}

\end{document}